\documentclass{elsart}
\usepackage{amsmath,amssymb} 
\usepackage{microtype}
\usepackage{tikz}

\usetikzlibrary{arrows}
\usetikzlibrary{automata}

\newcommand{\N}{\mathbb{N}}
\renewcommand{\P}{\mathbb{P}}
\newcommand{\T}{\mathcal{T}}
\newcommand{\pdl}{\textnormal{PDL}}
\newcommand{\PDL}{\pdl$(k,n)$}
\newcommand{\Tests}{\mathrm{Tests}}
\newcommand{\true}{\mathord{\mathtt{true}}}
\newcommand{\false}{\mathord{\mathtt{false}}}
\newcommand{\links}{[\![}
\newcommand{\rechts}{]\!]}
\renewcommand{\int}{\mathit{int}}
\newcommand{\snake}{\mathrm{snake}}
\newcommand{\sound}{\mathrm{tile}}
\newcommand{\recur}{\mathrm{recur}}
\newcommand{\diam}[1]{\mathord{\langle#1\rangle}}
\newcommand{\bokz}[1]{\mathord{\lbrack#1\rbrack}}

\begin{document}

\begin{frontmatter}

\title{On a Non-Context-Free Extension of PDL}
\author{Stefan G\"oller \and Dirk Nowotka}
\ead{\{goeller,nowotka\}@fmi.uni-stuttgart.de}
\address{Institute for Formal Methods in Computer Science (FMI) \\
  University of Stuttgart, Germany}

\begin{abstract}
  Over the last 25 years, a lot of work has been done on seeking
  for decidable non-regular extensions of Propositional Dynamic Logic (PDL).
  Only recently, an expressive extension of PDL, allowing visibly 
  pushdown automata (VPAs) as a formalism to describe programs,
  was introduced and proven to have a~satisfiability problem
  complete for deterministic double exponential time. 
  Lately, the VPA formalism was extended to so called $k$-phase 
  multi-stack visibly pushdown automata ($k$-MVPAs). Similarly to VPAs,
  it has been shown that the language of $k$-MVPAs have desirable 
  effective closure properties and that the emptiness problem is decidable. 
  On the occasion of introducing $k$-MVPAs, it has
  been asked whether the extension of PDL with $k$-MVPAs still leads
  to a decidable logic. This question is answered negatively here.
  We prove that already for the extension of PDL with $2$-phase MVPAs 
  with two stacks satisfiability becomes $\Sigma_1^1$-complete.
\end{abstract}

\end{frontmatter}

\section{Introduction}
Propositional Dynamic Logic (PDL) is a modal logic introduced by
Fischer and Ladner \cite{FiLa79} which allows to reason about regular programs.
In PDL, there are two syntactic entities: formulas, built from boolean
and modal operators and interpreted as sets of worlds of a Kripke structure;
and programs, built from the operators test, union, composition, and Kleene
star and interpreted as binary relations in a Kripke structure.
Thence, the occuring programs can be seen as a regular language over
an alphabet that consists of tests and atomic programs.
However, the mere usage of regular programs limits
the expressiveness of PDL as for example witnessed
by the set of executions of well-matched calls and returns of a recursive
procedure, cf. \cite{HKT2000}. Therefore, non-regular extensions of PDL
have been studied quite extensively
\cite{HKT2000,HaPnSt83,LoedingLutzSerreJLAP07,KorPnu84}.
An~extension of PDL
by a~class~$\mathcal{L}$ of languages means that in addition to regular
languages also languages in~$\mathcal{L}$ may occur in modalities
of formulas.

One interesting result on PDL extensions, among many others as summarized
in \cite{HKT2000},  is that already
the extension of PDL with the \emph{single} 
language $\{a^nba^n \mid n\geq 1\}$
leads to an undecidable logic \cite{HaPnSt83}.
In contrast to this negative result, Harel and Raz proved that adding
to PDL a \emph{single} language accepted by
a~single-minded pushdown automaton yields a decidable
logic \cite{HaRa93}. 
A simple-minded pushdown automaton is a restricted pushdown automaton,
where each input symbol determines the next control state,
the stack operation and the stack symbol to be pushed, in case
a push operation is performed.
Generalizing this concept, Alur and Madhusudan proposed
in~\cite{AlMa04} visibly pushdown languages which are
defined as languages accepted by visibly pushdown automata (VPAs).
A VPA is a pushdown automaton, where the stack operation
is determined by the input in the following way;
the alphabet is partitioned into letters that prompt a push,
internal, or pop action, respectively.
Note that it is well-known that visibly pushdown automata
are strictly more powerful than simple-minded pushdown automata. 
Recently, also for the model of visibly pushdown languages, a PDL 
extension has been investigated by L\"oding, Lutz, and Serre
\cite{LoedingLutzSerreJLAP07}.
They proved that satisfiability of this PDL extension is
complete for deterministic double exponential time.
Note that for this result, every visibly pushdown language occuring 
in a formula must be over the same partition of the alphabet.

Recently, $k$-phase multi-stack visibly pushdown automata ($k$-MVPAs),
a~natural extension of VPAs, have been introduced in \cite{LaMaPa07}.
A $k$-MVPA is an automaton equipped with $n$ stacks where, again,
the actions on the stacks are determined by the input, more precisely,
every input symbol specifies on which stack a~push or pop operation
or whether an internal operation is done. 
Moreover, a~$k$-MVPA is restricted to accept only words that can
be obtained by concatenating at most $k$ {\em phases}, where a phase is
a~sequence of input symbols that invoke pop actions from at most one stack.
Note that $k$-MVPAs 
with one stack coincide with VPAs.

Due to the various effective closure properties and a decidable
emptiness problem of the language class described by $k$-MVPAs, 
it is an interesting question to ask if 
the corresponding extension of PDL is still decidable.
This question 
was raised in \cite{LaMaPa07} and is answered negatively in this article.
We prove $\Sigma_1^1$-completeness for this PDL extension.
A $\Sigma_1^1$ lower bound already holds, if we restrict ourselves
to deterministic $2$-MVPAs with two stacks.
This is the weakest possible instance of $k$-MVPAs that is still
more powerful than VPAs. 
Our proof relies on the same technique of the $\Sigma_1^1$-hardness
proof of undecidability of PDL extended with the single language
$\{a^nba^n\mid n\geq 1\}$, which is presented
in \cite{HKT2000}. Note however, that $\{a^nba^n\mid n\geq 1\}$ is not
recognized by any $k$-MVPA for any $k$.

We proceed as follows.
We recapitulate $k$-MVPAs in Section \ref{S KMVPA}.
Section \ref{S Logic} introduces the extension of PDL with
$k$-MVPAs. 
A $\Sigma_1^1$-completeness proof is presented in 
Section \ref{S Undecidability}.

\section{$k$-Phase Multi-Stack Visibly Pushdown {Auto\-mata}}{\label{S KMVPA}}
In this section we recall the definition of $k$-phase multi-stack visibly
pushdown automata from \cite{LaMaPa07}.

Let $\N=\{0,1,2,\ldots\}$ denote the natural numbers.
Let $n\in\N$, then $[n]=\{1,2,\ldots,n\}$. 
Note that $[0]=\emptyset$.
Let $\varepsilon$ denote the {\em empty word}.
For some $n\in\N$ an {\em $n$-stack call-return alphabet} is a tuple
$\widetilde{\Sigma}_n=\langle\{\Sigma^i_c,\Sigma^i_r\}_{i\in[n]},\Sigma_\int\rangle$ of pairwise disjoint finite alphabets. 
Let $\Sigma_c=\bigcup_{i\in[n]}\Sigma^i_c$ and
$\Sigma_r=\bigcup_{i\in[n]}\Sigma^i_r$ for every
$i\in[n]$, and let $\Sigma=\Sigma_c\cup\Sigma_r\cup\Sigma_\int$.
Let us fix $\widetilde{\Sigma}_n$ for the rest of this section.

\begin{defn}
A multi-stack visibly pushdown automaton (MVPA) over
$\widetilde{\Sigma}_n$ is a tuple
$M=(Q,Q_I,\Gamma,\delta,Q_F)$, where
(i) $Q$ is a finite set of {\em states},
(ii) $Q_I\subseteq Q$ is the set of {\em initial states},
(iii) $\Gamma$ is a finite {\em stack alphabet} with
$\bot\in\Gamma\setminus\Sigma$,
(iv) $\delta\subseteq (Q\times\Sigma_c\times Q\times
\Gamma\setminus\{\bot\})\cup (Q\times\Sigma_r\times\Gamma\times Q)
\cup (Q\times\Sigma_\int\times Q)$,
and (v) $Q_F\subseteq Q$ is the set of {\em final states}.
\end{defn}
A $k$-MVPA is {\em deterministic}, if 
$|Q_I|=1$ and
for each $q\in Q$, for each $a\in\Sigma$, and for each $\gamma\in\Gamma$ we
have
$$\left|\delta\ \cap\ 
\left(\{q\}\times\{a\}\times (Q\times\Gamma\setminus\{\bot\}\cup
 \{\gamma\}\times Q\cup  Q)\biggl.\right)\right|\leq 1.$$
The set of {\em stacks} is defined as
$St=(\Gamma\setminus\{\bot\})^*\cdot\{\bot\}$. 
A {\em configuration} of an MVPA is a pair $(q,C)$ where
$q\in Q$ and $C:[n]\rightarrow St$ is a mapping.
A {\em run} of $M$ on an input $w=a_1a_2\cdots a_m\in\Sigma^* (m\geq 0)$, 
  with $a_i\in\Sigma$ for each $i\in[m]$, is a sequence of configurations
$(q_0,C_0)(q_1,C_1)\cdots(q_m,C_m)$  such that 
\begin{itemize}
\item $q_0\in Q_I$ and $C_0(i)=\bot$ for each $i\in[n]$ and
\item for every $j\geq 1$ we have,
\begin{itemize}
\item whenever $a_j\in\Sigma^i_c$ for some $i\in[n]$, then
there exists some $\gamma\in\Gamma\setminus\{\bot\}$ such that
$(q_{j-1},a_j,q_j,\gamma)\in\delta$, and $C_j(i)=\gamma\cdot C_{j-1}(i)$
and $C_j(i')=C_{j-1}(i')$ for all $i'\in[n]$ with $i'\not=i$,
\item whenever $a_j\in\Sigma^i_r$ for some $i\in[n]$, then
there exists some $\gamma\in\Gamma$ such that
$(q_{j-1},a_j,\gamma,q_j)\in\delta$, and $C_j(i')=C_{j-1}(i')$ for
all $i'\in[n]$ with $i'\not=i$ and either 
(i) $\gamma=\bot$ and $C_j(i)=C_{j-1}(i)=\bot$ or 
(ii) $\gamma\not=\bot$ and
$C_{j-1}(i)=\gamma\cdot C_j(i)$, and 
\item whenever $a_j\in\Sigma_\int$, then 
$(q_{j-1},a_j,q_j)\in\delta$ and $C_j(i)=C_{j-1}(i)$ for all $i\in[n]$.
\end{itemize}
\end{itemize}
We call a run $(q_0,C_0)(q_1,C_1)\cdots (q_m,C_m)$ {\em accepting}, if
$q_m\in Q_F$. 
Furthermore, we denote
by $L(M)=\{w\in\Sigma^*\mid\text{ there exists an accepting run of }
M\text{ on }w\}$ the {\em language} of $M$.
A~word $w\in\Sigma^*$ is a {\em phase}, if 
$w\in(\Sigma_c\cup\Sigma_\int\cup\Sigma^i_r)^*$ for
some $i\in\N$. For $k\geq 1$, we say a word is a {\em $k$-phase}
if it can be obtained by~concatenating at most $k$ phases.
\begin{defn}
  A {\em $k$-phase multi-stack visibly pushdown automaton} ($k$-MVPA)
  $M$ is a~{mul\-ti}-stack visibly pushdown automaton that is restricted to 
  accept $k$-phases only. Formally, we define
  $$L(M)=\{w\in\Sigma^*\mid w\mbox{\ is\ a\ } k\mbox{-phase
  and\ there\ exists\ an\ accepting\ run\ of\ }M\mbox{\ on\ }w\}.$$
\end{defn}
Note that $n=0$ implies that a $k$-MVPA is as powerful as a finite
state automaton. Moreover, we get precisely the VPAs as introduced
in~\cite{AlMa04} when $n=1$.

\section{Propositional Dynamic Logic over $k$-MVPAs}{\label{S Logic}}
Fix some countable set $\P$ of {\em atomic propositions}, and some
$k,n\in\N$ with $k\geq 1$.
The {\em set of formulas} $\Phi$ and the set of {\em tests} $\Tests$
of the logic \PDL\ over some $n$-stack call-return alphabet 
$\widetilde{\Sigma}_n=\langle\{\Sigma^i_c,\Sigma^i_r\}_{i\in[n]},
  \Sigma_\int\rangle$  
  are the smallest sets that
  satisfy the following conditions:
  \begin{itemize}
  \item $\true\in\Phi$,
  \item if $p\in\P$, then $p\in\Phi$,
  \item if $\varphi_1,\varphi_2\in\Phi$, then 
  $\varphi_1\vee\varphi_2,\neg\varphi_1\in\Phi$,
  \item if $\varphi\in\Phi$, then $\varphi?\in\Tests$
  \item if $\varphi\in\Phi$ and $\Psi\subset\Tests$ is finite, 
  then $\langle \chi\rangle\varphi\in\Phi$, where
  $\chi$ is either a regular expression over $\Sigma\cup\Psi$ or
  $\chi$ is a $k$-MVPA over
  $\langle\{\Sigma^i_c,\Sigma^i_r\}_{i\in[n]},\Sigma_\int\cup\Psi\rangle$.
  \end{itemize}
We introduce the usual abbreviations $\false=\neg\true$, 
$\varphi_1\wedge\varphi_2=\neg(\neg\varphi_1\vee\neg\varphi_2)$,
and $\lbrack \chi\rbrack\varphi=\neg\langle \chi\rangle\neg\varphi$.
A {\em Kripke structure} is a tuple 
$K=(X,\{\rightarrow_a\}_{a\in\Sigma},\rho)$, where $X$ is a set of 
{\em worlds}, $\rightarrow_a\subseteq X\times X$ is a binary relation
for each $a\in\Sigma$,
and $\rho:X\rightarrow 2^\P$ assigns to each world a set of atomic
propositions.
For each $\varphi\in\Phi$ and for each $w\in(\Sigma\cup\Tests)^*$, define
the binary relation
$\links w\rechts_K\subseteq X\times X$ and the set 
$\links\varphi\rechts_K\subseteq X$ via mutual induction as follows:
\begin{itemize}
\item $\links\varepsilon\rechts_K=\{(x,x)\mid x\in X\}$,
\item if $\varphi?\in\Tests$, then
$\links\varphi?\rechts_K=\{(x,x)\mid x\in X\wedge 
x\in\links\varphi\rechts_K\}$,
\item if $a\in\Sigma$, then $\links a\rechts_K=\ \rightarrow_a$,
\item if $w\in(\Sigma\cup\Tests)^*$ and $\tau\in\Sigma\cup\Tests$, then
$\links w\tau\rechts_K=\links w\rechts_K\circ\links \tau\rechts_K$,
\item if $p\in\P$, then $\links p\rechts_K=\{x\in X\mid p\in\rho(x)\}$,
\item $\links\varphi_1\vee\varphi_2\rechts_K=
\links\varphi_1\rechts_K\cup\links\varphi_2\rechts_K$,
\item $\links\neg\varphi\rechts_K=X\setminus\links\varphi\rechts_K$,
\item $\links\langle \chi\rangle\varphi\rechts_K=
\{x\in X\mid \exists y\in X\ \exists w\in L(\chi): 
  (x,y)\in\links w\rechts_K\wedge y\in\links\varphi\rechts_K\}$.
\end{itemize}
Note that since we restrict $k$-MVPAs to accept $k$-phases only,
we additionally allow formulas of the kind 
$\langle\alpha\rangle\varphi$, where $\alpha$ is a regular expression
over a finite subset of $\Sigma\cup\Tests$.
A $k$-MVPA can accept a regular language over $k$-phases only, that is,
not even $\Sigma^*$ (if $\Sigma$ contains two pop symbols from
different stacks) can be recognized.
However, since we would like to increase the expressiveness of PDL
beyond regular programs, we have to explicitly take in regular
expressions. 
If $L$ is a language over a finite subset of $\Sigma\cup\Tests$, we 
define $\links L\rechts_K=\bigcup_{w\in L}\links w\rechts_K.$
In the following, we will write $\langle L \rangle\varphi$ 
($\lbrack L\rbrack\varphi$)
instead $\langle\chi\rangle\varphi$ ($\lbrack\chi\rbrack\varphi)$, where
$L$ is the language of $\chi$ and $\chi$ is either 
some regular expression or
some $k$-MVPA.
We also write $(K,x)\models\varphi$ whenever $x\in\links\varphi\rechts_K$.
We say that $K$ is a {\em model} for $\varphi$, if $(K,x)\models\varphi$
for some world $x$ of $K$. We say a \PDL\ formula $\varphi$ is 
{\em satisfiable}, if there exists a model for $\varphi$.
The {\em satisfiability problem} asks, given a \PDL\ formula $\varphi$,
  whether $\varphi$ is satisfiable.

When restricting all automata that occur in a formula to be
visibly pushdown automata (i.e. over a single stack), 
L\"oding, Lutz and Serre  obtained the following result:

\begin{thm}[\cite{LoedingLutzSerreJLAP07}]
Satisfiability of \textnormal{PDL}$(1,1)$ is
complete for deterministic double exponential time.
\end{thm}

\section{$\Sigma_1^1$-Completeness of PDL$(k,n)$}{\label{S Undecidability}}
For the $\Sigma^1_1$ upper bound,
we can easily adapt the proof of Proposition 9.4 
in~\cite{HKT2000} and show that every satisfiable PDL$(k,n)$ formula 
has a countable tree model. Thus, we can write down an existential 
second-order number-theoretic formula over $\N$ that is valid 
if and only if $\varphi$ is satisfiable.

For the lower bound, we prove that PDL$(k,n)$
is $\Sigma_1^1$-hard already for $k=2$ and $n=2$,
i.e. we can restrict all occurring MVPAs to have $2$~stacks and
to accept $2$-phases only.
For this, we reduce the $\Sigma_1^1$-hard recurring tiling problem of the 
first quadrant of the plane to satisfiability
of PDL$(2,2)$. 
A {\em recurring tiling system}  $\T=(T,H,V,t_0)$ consists of a 
finite set of 
{\em tile types} $T$, a {\em horizontal matching relation}
$H\subseteq T\times T$, a {\em vertical matching relation}
$V\subseteq T\times T$, and a~tile type $t_0\in T$.
A {\em solution for~$\T$}
is a mapping $\mu:\N\times\N\rightarrow T$
such that for infinitely many $m\in\N$ we have $\mu(0,m)=t_0$ and
for all $(n,m)\in\N\times\N$ we have
\begin{itemize}
  \item if $\mu(n,m)=t$ and $\mu(n+1,m)=t'$, then $(t,t')\in H$, and 
  \item if $\mu(n,m)=t$ and $\mu(n,m+1)=t'$, then $(t,t')\in V$.
\end{itemize}
The {\em recurring tiling problem} is to decide whether a given 
recurring tiling system has a~solution.
\begin{thm}[\cite{Har84}]
The recurring tiling problem is $\Sigma_1^1$-complete.
\end{thm}
For the rest of the section fix some tiling system
$\T=(T,H,V,t_0)$. Our goal is to translate $\T$ into a 
PDL$(2,2)$ formula $\varphi=\varphi(\T)$ over 
the set of atomic propositions $T$ such that
$\T$ has a solution if and only if $\varphi$ is satisfiable.

Fix the $2$-stack alphabet
$\widetilde{\Sigma}_2=
\langle\{\Sigma^i_c,\Sigma^i_r\}_{i\in\{1,2\}},\Sigma_\int\rangle$
where $\Sigma^i_c=\{a_i\}$ and $\Sigma^i_r=\{b_i\}$ for each 
$i\in[2]$ and where $\Sigma_\int=\{c,d\}$.
Define the languages $L_\ell$, $L_\ell^\leftrightarrow$,
and $L_\ell^\updownarrow$ for each
$\ell\in\{0, 1\}$ as follows,
where $w_0=a_1b_2$ and $w_1=a_2b_1$ and $e_0=d$ and $e_1=c$:
\begin{eqnarray*}
  L_\ell &\qquad=\qquad &\{w_\ell^i\,e_\ell\,w_{1-\ell}^j\,e_{1-\ell}
    \mid i,j\geq 0 \text{ and } j\neq i+1\} , \\
  L_\ell^\leftrightarrow &\qquad=\qquad
    &\{w_\ell^i\,e_\ell\,w_{1-\ell}^{i+\ell+1}
    \mid i\geq 0\} , \\
  L_\ell^\updownarrow &\qquad=\qquad
    &\{w_\ell^i\,e_\ell\,w_{1-\ell}^{i-\ell+2}
    \mid i\geq 0\} .
\end{eqnarray*}

\begin{prop}{\label{P Languages}}
For each of the languages $L_\ell$, $L_\ell^\leftrightarrow$,
and $L_\ell^\updownarrow$, with $\ell\in\{0, 1\}$,
there exists a deterministic $2$-MVPA over $\widetilde{\Sigma}_2$
that accepts it.
\end{prop}
\begin{pf}
  Figures \ref{fig:Mplus} to \ref{fig:Mns} depict $2$-MVPAs recognizing
  $L_\ell$, $L_\ell^\leftrightarrow$,
  and $L_\ell^\updownarrow$, respectively, for $\ell=0$.
  The case $\ell=1$ is deduced by simultaneously
  substituting $a_1$, $b_2$, $c$, and $d$
  by~$a_2$, $b_1$, $d$, and~$c$, respectively.
  Note that all automata are deterministic.
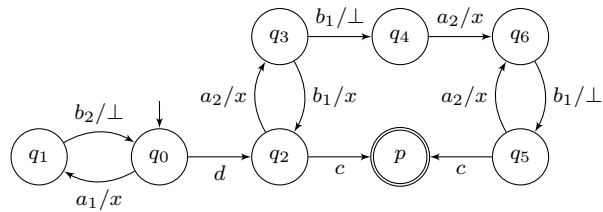
\begin{figure}[ht]
{\scriptsize
\begin{center}
\begin{tikzpicture}
  [node distance=1.6cm,auto,>=latex',initial text=,initial where=above]
  \node[state,initial] (q_0) {$q_0$};
  \node[state] (q_1) [left of=q_0] {$q_1$};
  \node[state] (q_2) [right of=q_0] {$q_2$}; 
  \node[state] (q_3) [above of=q_2] {$q_3$};
  \node[state] (q_4) [right of=q_3] {$q_4$}; 
  \node[state,accepting] (p) [right of=q_2] {$p$};
  \node[state] (q_5) [right of=p] {$q_5$}; 
  \node[state] (q_6) [right of=q_4] {$q_6$}; 
  \path[->] (q_0) edge [bend left] node {$a_1 / x$} (q_1)
            (q_1) edge [bend left] node {$b_2 / \bot$} (q_0)
            (q_0) edge node [swap] {$d$} (q_2)
            (q_2) edge [bend left] node {$a_2 / x$} (q_3)
            (q_3) edge [bend left] node {$b_1 / x$} (q_2)
            (q_3) edge node {$b_1 / \bot$} (q_4)
            (q_4) edge node {$a_2 / x$} (q_6)
            (q_5) edge [bend left] node {$a_2 / x$} (q_6)
            (q_6) edge [bend left] node {$b_1 / \bot$} (q_5)
            (q_5) edge node {$c$} (p)
            (q_2) edge node [swap] {$c$} (p);
\end{tikzpicture}
\end{center}
}
\caption{A $2$-MVPA recognizing
  $L_0=\{(a_1 b_2)^i d (a_2 b_1)^j c \mid i, j\geq 0, 
  j\neq i+1\}$.}
\label{fig:Mplus}
\end{figure}
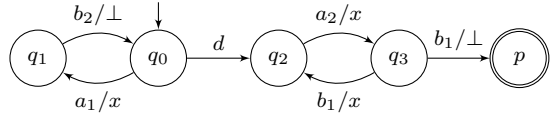
\begin{figure}[ht]
{\scriptsize
\begin{center}
\begin{tikzpicture}
  [node distance=1.6cm,auto,>=latex',initial text=,initial where=above]
  \node[state,initial] (q_0) {$q_0$};
  \node[state] (q_1) [left of=q_0] {$q_1$};
  \node[state] (q_2) [right of=q_0] {$q_2$}; 
  \node[state] (q_3) [right of=q_2] {$q_3$};
  \node[state,accepting] (p) [right of=q_3] {$p$}; 
  \path[->] (q_0) edge node {$d$} (q_2)
            (q_0) edge [bend left] node {$a_1 / x$} (q_1)
            (q_1) edge [bend left] node {$b_2 / \bot$} (q_0)
            (q_2) edge [bend left] node {$a_2 / x$} (q_3)
            (q_3) edge [bend left] node {$b_1 / x$} (q_2)
            (q_3) edge node {$b_1 / \bot$} (p);
\end{tikzpicture}
\end{center}
}
\caption{A $2$-MVPA recognizing
  $L_0^\leftrightarrow=\{(a_1 b_2)^i d (a_2 b_1)^{i+1} \mid i\geq 0\}$.}
\label{fig:Mew}
\end{figure}
\begin{figure}[ht]
{\scriptsize
\begin{center}
\begin{tikzpicture}
  [node distance=1.6cm,auto,>=latex',initial text=,initial where=above]
  \node[state,initial] (q_0) {$q_0$};
  \node[state] (q_1) [left of=q_0] {$q_1$};
  \node[state] (q_2) [right of=q_0] {$q_2$}; 
  \node[state] (q_3) [right of=q_2] {$q_3$};
  \node[state] (q_4) [right of=q_3] {$q_4$};
  \node[state] (q_5) [right of=q_4] {$q_5$};
  \node[state,accepting] (p) [right of=q_5] {$p$}; 
  \path[->] (q_0) edge node {$d$} (q_2)
            (q_0) edge [bend left] node {$a_1 / x$} (q_1)
            (q_1) edge [bend left] node {$b_2 / \bot$} (q_0)
            (q_2) edge [bend left] node {$a_2 / x$} (q_3)
            (q_3) edge [bend left] node {$b_1 / x$} (q_2)
            (q_3) edge node {$b_1 / \bot$} (q_4)
            (q_4) edge node {$a_2 / x$} (q_5)
            (q_5) edge node {$b_1 / \bot$} (p);
\end{tikzpicture}
\end{center}
}
\caption{An automaton recognizing
  $L_0^\updownarrow=\{(a_1 b_2)^i d (a_2 b_1)^{i+2} \mid i\geq 0\}$.}
\label{fig:Mns}
\end{figure}
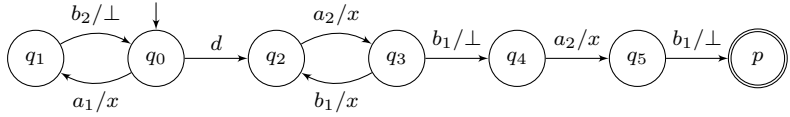
\qed\end{pf}

Let $\varphi_\snake$ be defined as follows:
\begin{align*}
  \varphi_\snake \qquad=\qquad 
  \diam{ca_1b_2d(a_2b_1)^2c}\true\ &\land\ %
    \bokz{\Sigma^*c}(\diam{(a_1b_2)^* d}\true \land \bokz{L_0}\false) \\
    &\land\ \bokz{\Sigma^*d}(\diam{(a_2b_1)^* c}\true \land \bokz{L_1}\false).
\end{align*}
A {\em snake} of a Kripke structure $K$ is an infinite path in $K$
that is labeled~by
\[
  c(a_1b_2)^1 d (a_2b_1)^2 c (a_1b_2)^3 d (a_2b_1)^4
  c(a_1b_2)^5 d (a_2b_1)^6 c \cdots.
\]

\begin{prop}{\label{P Infinite Path}}
  Every model of $\varphi_\snake$ has a snake.
\end{prop}
\begin{pf}
  Let $K=(X,\{\rightarrow_a\}_{a\in\Sigma},\rho)$ be a model of 
  $\varphi_\snake$, i.e. $(K,x)\models\varphi_\snake$ for 
  some $x\in X$.
  By the first conjunct of $\varphi_\snake$, there exist
  worlds $x_1,x_2\in X$ such that
  $(x,x_1)\in\links ca_1b_2d\rechts_K$, and 
  $(x_1,x_2)\in\links (a_2b_1)^2c\rechts_K$.
  Firstly, observe that $(K,x_2)\models\diam{(a_1b_2)^+ d}\true$
  by the third conjunct of $\varphi_\snake$.
  This implies that
  $(x_2,x_3)\in\links (a_1b_2)^{i}d\rechts_K$ for
  some $x_3\in X$ and some $i\in\N$.
  But clearly $i=3$, for otherwise 
  $(K,x_1)\not\models\lbrack L_1\rbrack\false$.
  Thus we get
  $(x_2,x_3)\in\links (a_1b_2)^3d\rechts_K$.
  Symmetrically, since
  $(K,x_3)\models\diam{(a_2b_1)^+ c}\true$
  and $(K,x_2)\models\bokz{L_0}\false$ by the second
  conjunct of $\varphi_\snake$,  
  there exists a world $x_4$ such that
  $(x_3,x_4)\in\links (a_2b_1)^4c\rechts_K$.  
  By repeatedly applying the above argument, it is straightforward to see
  that there exists an infinite sequence of worlds
  $x_1, x_2, x_3, x_4, \ldots$ such that for each $i\geq 1$ we have
  $(x_{2i-1},x_{2i})\in\links(a_2b_1)^{2i}c\rechts_K$
  and also
  $(x_{2i},x_{2i+1})\in\links(a_1b_2)^{2i+1} d\rechts_K$.
  Since additionally we have
  $(x,x_1)\in\links c(a_1b_2)^1d\rechts$, there exists a~snake 
  in $K$.
\qed\end{pf}
Let the programs $\pi_{\downarrow\uparrow}$ and $\beta$ and
the formula $\varphi_\recur$ be defined as follows:
\begin{eqnarray*}
  \pi_{\downarrow\uparrow} & \qquad=\qquad & (a_1b_2)^*d(a_2b_1)^*c\ , \\
  \beta &\qquad=\qquad &\pi_{\downarrow\uparrow}^*\ \left( a_1 b_2(t_0?)
    \pi_{\downarrow\uparrow}
    \cup (a_1b_2)^*d(a_2b_1)^*(t_0?)c\biggl.\right), \\
  \varphi_\recur &=& \bokz{\Sigma^*c}\diam{\beta}\true .
\end{eqnarray*}
We call a world $y$ on a snake $\sigma$ {\em first column}, if
either $x_1\xrightarrow{c}x_2\xrightarrow{a_1}x_3\xrightarrow{b_2}y$
or $y\xrightarrow{c}x$ is a subpath of $\sigma$.

\begin{prop}{\label{P First Column}}
  Every model of $\varphi_\snake\wedge\varphi_\recur$ has a snake 
  on which infinitely often first column worlds satisfy the atomic 
  proposition $t_0$.
\end{prop}
\begin{pf}
Let $K$ be a model of $\varphi_\snake\wedge\varphi_\recur$.
By Proposition \ref{P Infinite Path}, there exists a~snake $\sigma_0$
in $K$.  Fix an arbitrary world $x_0$ on $\sigma_0$ such that 
for some $x\in X$ we have that
$x\xrightarrow{c}_Kx_0$ is a subpath of $\sigma_0$.
It is not hard to see that, by definition of $\varphi_\recur$ and by 
similar arguments as in the proof
of Proposition \ref{P Infinite Path}, there exists a snake $\sigma_1$
whose initial part agrees with $\sigma_0$ up to world $x_0$ and such
that for some world $x_0'$ on $\sigma_1$, we have
$(x_0,x_0')\in\links\beta\rechts_K$. 
Moreover, by definition of $\beta$, on the subpath of $\sigma_1$
from $x_0$ to $x_0'$ there exists some first column world 
that satisfies the atomic proposition $t_0$.
Fix an arbitrary world $x_1$ on $\sigma_1$ 
such that there is a subpath from $x_0'$ to $x_1$ on $\sigma_1$
such that additionally 
for some $x'\in X$ we have that
$x'\xrightarrow{c}_Kx_1$ is a subpath
of $\sigma_1$. 
Again, we have $(K,x_1)\models\langle\beta\rangle\true$.
Hence again, there exists some snake $\sigma_2$ whose
initial part agrees with $\sigma_1$ up to $x_1$ such that for some
world $x_1'$ on $\sigma_2$ we have $(x_1,x_1')\in\links\beta\rechts_K$
and on the subpath of $\sigma_2$ from $x_1$ to $x_1'$ some first
column world of $\sigma_2$ satisfies $t_0$. 
By repeatedly applying the same argument, we obtain
a~snake in $K$ on which infinitely often first column worlds satisfy
$t_0$.
\qed\end{pf}

Let us now give a formula $\varphi_\sound$ that guarantees that
every (reachable) world contains exactly one tile type:
\[
  \varphi_\sound\qquad=\qquad\bokz{\Sigma^*}
  \left(\bigvee_{t\in T}\left(t\wedge\bigwedge_{t'\in T:t\not=t'}
  \neg(t\wedge t')\right)\right)
\]
Next, we give a formula $\varphi_\leftrightarrow^\updownarrow$ that 
ensures that the types of vertically (horizontally) connected tiles
satisfy the vertical (horizontal) matching relation:
\begin{align*}
  \varphi_\leftrightarrow^\updownarrow\ =\ %
  &\bokz{\Sigma^*c(a_1b_2)^+}
  \bigwedge_{t\in T}t\rightarrow
  \left(
  \bokz{L_0^\leftrightarrow} \bigvee_{t'\in T:(t,t')\in H}t'\ 
  \wedge\ \bokz{L_0^\updownarrow}\bigvee_{t'\in T:(t,t')\in V}t'\right)
  \ \land \\
  &\bokz{\Sigma^*d(a_2b_1)^+}
  \bigwedge_{t\in T}t\rightarrow
  \left(
  \bokz{L_1^\leftrightarrow} \bigvee_{t'\in T:(t,t')\in H}t'\ 
  \wedge\ \bokz{L_1^\updownarrow}\bigvee_{t'\in T:(t,t')\in V}t'\right)
\end{align*}
Our final formula $\varphi$ is 
\[
  \varphi\qquad=\qquad\varphi_\snake\wedge\varphi_\recur\wedge
  \varphi_\sound\wedge
  \varphi^\updownarrow_\leftrightarrow.
\]
Before proving that $\T$ has a solution if and only if $\varphi$ is
satisfiable, we introduce some more notation.
Let $A=\{(i,j)\in\N\times\N\mid 0\leq j\leq i\}$.
We define a bijection $\pi:\N\times\N\rightarrow A$ for all
$(n,m)\in\N\times\N$ as follows
\[
  \pi(n,m)\qquad =\qquad (n+m,m).
\]
Thus, $\pi^{-1}(i,j)=(i-j,j)$ for all $(i,j)\in A$.

\begin{lem}{\label{L Equivalence}}
  The recurring tiling system $\T$ has a solution if and only 
  if $\varphi$ is satisfiable.
\end{lem}
\begin{pf}\ \\
  \emph{only-if:} Assume that $\T$ has a solution $\mu:\N\times\N\rightarrow T$.
  Figure \ref{F Model} depicts a~model
  $K=K(\T)=(X,\{\rightarrow_a\}_{a\in\Sigma},\rho)$
  that we can construct from $\T$.
  To all those worlds that are pictured by bullets, the mapping $\rho$
  assigns an arbitrary singleton subset from $T$.
  For the worlds $x_{i,j}$, where $(i,j)\in A$, we define 
  \[
    \rho(x_{i,j})=\mu(\pi^{-1}(i,j)).
  \]
  Thus, the world $x_{i,j}$ represents the unique  
  the pair $(n,m)\in\N$ such that $\pi(n,m)=(i,j)$. 
  It is straightforward to verify that $(K,x)\models\varphi$.
\setlength{\unitlength}{0.05cm}
\begin{center}
\begin{figure}[htb]
{\scriptsize
\begin{center}
\begin{tikzpicture}
  [>=latex',step=2cm]
  \draw (0,0) grid (10,10);
  \node[draw] (n) at (-1cm, 1cm) {$x$};
  \draw[->,shorten >=2pt] (n) -- node[right] {$c$} (-1cm, 3cm);
  \foreach \x in {0, 2, ..., 4}
  {
    \foreach \y in {0, ..., \x}
    {
      \node[draw] (n)
        at (2cm*\x - 2cm*\y + 1cm, 2cm*\y + 1cm) {$x_{\x,\y}$};
      \draw[fill] (2cm*\x - 2cm*\y, 2cm*\y + 2cm) circle (2pt);
      \draw[->,shorten <=2pt] (2cm*\x - 2cm*\y, 2cm*\y + 2cm) -- 
        node[above right=-2pt] {$b_2$} (n);
      \node (m)
        at (2cm*\x - 2cm*\y - 1cm, 2cm*\y + 3cm) {\phantom{$x_{\x,\y}$}};
      \ifnum \y<\x
        {
          \draw[->,shorten >=2pt]
            (m) -- node[above right=-2pt] {$a_1$}
              (2cm*\x - 2cm*\y, 2cm*\y + 2cm);
        }
      \fi
    }
    \draw[->,shorten <=2pt,shorten >=2pt]
      (-1cm,2cm*\x + 3cm) -- node[above right=-2pt] {$a_1$}
        (0cm, 2cm*\x + 2cm);
    \draw[fill] (-1cm,2cm*\x + 3cm) circle (2pt);
  };
  \foreach \x in {1, 3}
  {
    \node (m) at (2cm*\x - 1cm, 1cm) {\phantom{$x_{\x,\x}$}};
    \draw[->,shorten >=2pt] (m)
      .. controls +(315:3cm) and +(180:2cm) ..
      node[above right=-2pt] {$d$}
      (2cm*\x + 3cm, -1cm);
    \draw[fill] (2cm*\x + 3cm, -1cm) circle (2pt);
    \draw[->,shorten <=2pt,shorten >=2pt]
      (2cm*\x + 3cm, -1cm) -- node[above right=-2pt] {$a_2$}
        (2cm*\x + 2cm, 0);
    \foreach \y in {0, ..., \x}
    {
      \node[draw] (n)
        at (2cm*\x - 2cm*\y + 1cm, 2cm*\y + 1cm) {$x_{\x,\y}$};
      \draw[fill] (2cm*\x - 2cm*\y + 2cm, 2cm*\y) circle (2pt);
      \draw[->,shorten <=2pt] (2cm*\x - 2cm*\y + 2cm, 2cm*\y) -- 
        node[above right=-2pt] {$b_1$} (n);
      \node (m)
        at (2cm*\x - 2cm*\y + 3cm, 2cm*\y - 1cm) {\phantom{$x_{\x,\y}$}};
      \ifnum \y>0
        {
          \draw[->,shorten >=2pt] (m) --
            node[above right=-2pt] {$a_2$} (2cm*\x - 2cm*\y + 2cm, 2cm*\y);
        }
      \fi
    }
    \draw[->,shorten >=2pt] (n)
      .. controls +(135:3cm) and +(270:2cm) ..
      node[above right=-2pt] {$c$} (-1cm, 2cm*\x + 5cm);
  };
  \node (n) at (9cm, 1cm) {\phantom{$x_{4, 0}$}};
  \draw[->,shorten >=5pt] (n) -- node[above right=-2pt] {$d$} (10cm, 0cm);
  \path (10cm, 0.2cm) node[below right] {$\ddots$};
\end{tikzpicture}
\end{center}
}
\caption{Constructing a model from a solution of $\T$.}
{\label{F Model}}
\end{figure}
\end{center}
  \emph{if:} Let $K=(X,\{\rightarrow_a\}_{a\in\Sigma},\rho)$ be a
  model of $\varphi$, i.e. we have
  $(K,x)\models\varphi$ for some world $x\in X$ . 
  We prove that $\T$ has a solution.
  By Proposition \ref{P First Column} there exists a~snake
  $\sigma$ in $K$ on which infinitely often first column worlds satisfy
  the atomic proposition $t_0$, since both
  $\varphi_\snake$ as well as $\varphi_\recur$ occur in $\varphi$ 
  as a conjunct and $K$ is a model of $\varphi$.
  Recall that $A=\{(i,j)\in\N\times\N\mid 0\leq j\leq i\}$.
  For each $(i,j)\in A$, fix some world $x_{i,j}$ on $\sigma$
  such that $x\xrightarrow{ca_1 b_2}_Kx_{0,0}$ and the following holds
  for each $r\in\N$: 
  \[
    x_{2r,0}\xrightarrow{da_2b_1}_K x_{2r+1,0}
    \qquad\text{ and }\qquad
    x_{2r,s}\xrightarrow{a_1b_2}_K x_{2r,s-1}
  \]
  and
  \[
    x_{2r+1, 2r+1}\xrightarrow{ca_1b_2}_K x_{2r+2, 2r+2}
    \qquad\text{ and }\qquad
    x_{2r+1,s}\xrightarrow{a_2b_1}_K x_{2r+1, s+1}
  \]
  for all $0\leq s \leq 2r$.
  Note that the first column nodes of $\sigma$ are precisely the
  nodes $\{x_{m,m}\mid m\in\N\}$.
  Moreover, for all $(2r,s), (2r+1,s)\in A$ we have
  \begin{eqnarray}
    (x_{2r,s},x_{2r+1,s})&\ \in\ %
      &\links (a_1b_2)^{2r-s} d (a_2b_1)^{2r-s+1}\rechts_K ,
      \label{E A}\\
    (x_{2r+1,s},x_{2r+2,s})&\ \in\ %
      &\links (a_2b_1)^{2r-s+1} c (a_1b_2)^{2r-s+3}\rechts_K , \label{E B} \\
    (x_{2r,s},x_{2r+1,s+1})&\ \in\ %
      &\links (a_1b_2)^{2r-s} d (a_2b_1)^{2r-s+2}\rechts_K , \label{E C} \\
    (x_{2r+1,s},x_{2r+2,s+1})&\ \in\ %
      &\links (a_2b_1)^{2r-s+1} c (a_1b_2)^{2r-s+2}\rechts_K.
\label{E D}
  \end{eqnarray}
  Recall that
  \begin{eqnarray*}
    L_0^\leftrightarrow &\qquad=\qquad
      &\{(a_1b_2)^t\,d\,(a_2b_1)^{t+1} \mid t\geq 0\}
        \text{,} \\
    L_1^\leftrightarrow &\qquad=\qquad
      &\{(a_2b_1)^t\,c\,(a_1b_2)^{t+2} \mid t\geq 0\}
        \text{,} \\
    L_0^\updownarrow &\qquad=\qquad
      &\{(a_1b_2)^t\,d\,(a_2b_1)^{t+2} \mid t\geq 0\}
        \text{,} \\ 
    L_1^\updownarrow &\qquad=\qquad
      &\{(a_2b_1)^t\,c\,(a_1b_2)^{t+1} \mid t\geq 0\} .
  \end{eqnarray*}
  Summarizing (\ref{E A}) and (\ref{E B}), we obtain
  for all $(2r,s),(2r+1,s)\in A$
  \begin{eqnarray}
    (x_{2r,s},x_{2r+1,s})&\qquad \in\qquad %
      &\links L_0^{\leftrightarrow}\rechts_K , {\label{E Eastwest}}\\
    (x_{2r+1,s},x_{2r+2,s})&\qquad \in\qquad %
      &\links L_1^{\leftrightarrow}\rechts_K .
  \end{eqnarray} 
  Similarly, summarizing (\ref{E C}) and (\ref{E D}), we obtain
  for all $(2r,s),(2r+1,s)\in A$
  \begin{eqnarray}
    (x_{2r,s},x_{2r+1,s+1})&\qquad \in\qquad %
      &\links L_0^{\updownarrow}\rechts_K , {\label{E Northsouth}}\\
    (x_{2r+1,s},x_{2r+2,s+1})&\qquad \in\qquad %
      &\links L_1^{\updownarrow}\rechts_K.    
  \end{eqnarray}
  For the rest of the proof, we show that the following mapping
  $\mu\colon\N\times\N\to T$ is a solution for $\T$, where 
  $(n,m)\in\N\times\N$:
  \[
    \mu(n,m)=t\qquad\text{ if }\ \ \{t\}=\rho(x_{\pi(n,m)}).
  \]
  Note that $\mu$ is well-defined since the formula $\varphi_\sound$ 
  guarantees that $\rho(x_{\pi(n,m)})$ is indeed a singleton.
  Since each first column world on $\sigma$ is $x_{m,m}$ for some
  $m\in\N$, infinitely often first column worlds satisfy
  $t_0$, and $\pi^{-1}(m,m)=(0,m)$, it follows that $\mu(0,m)=t_0$ for
  infinitely many $m\in\N$.

  Fix some $(n,m)\in\N\times\N$ such that
  $n+m$ is even. The case when $n+m$ is odd can be handled 
  analogously.

   Let $\mu(n,m)=t$ and $\mu(n+1,m)=t'$ for some $t,t'\in T$.
  We prove that $(t,t')\in H$.
  By definition, we have $\rho(x_{\pi(n,m)})=\{t\}$ and
  $\rho(x_{\pi(n+1,m)})=\{t'\}$.
   
  Note that $\pi(n,m)=(n+m,m)$ and $\pi(n+1,m)=(n+m+1,m)$ and since
  $n+m$ is even, it follows by~(\ref{E Eastwest}) that
  \begin{eqnarray}
    (x_{\pi(n,m)},x_{\pi(n+1,m)})\qquad\in\qquad
    \links L_0^\leftrightarrow\rechts_K.
    {\label{E Path}}
  \end{eqnarray}
  Recall that $\varphi_\leftrightarrow^\updownarrow$ is defined as follows:
  \begin{align*}
    \varphi_\leftrightarrow^\updownarrow\ =\ %
    &\bokz{\Sigma^*c(a_1b_2)^+}
    \bigwedge_{t\in T}t\rightarrow
    \left(
    \bokz{L_0^\leftrightarrow} \bigvee_{t'\in T:(t,t')\in H}t'\ 
    \wedge\ \bokz{L_0^\updownarrow}\bigvee_{t'\in T:(t,t')\in V}t'\right)
    \ \land \\
    &\bokz{\Sigma^*d(a_2b_1)^+}
    \bigwedge_{t\in T}t\rightarrow
    \left(
    \bokz{L_1^\leftrightarrow} \bigvee_{t'\in T:(t,t')\in H}t'\ 
    \wedge\ \bokz{L_1^\updownarrow}\bigvee_{t'\in T:(t,t')\in V}t'\right)
  \end{align*}
  By 
  $(x,x_{\pi(n,m)})\in\links\Sigma^*c(a_1b_2)^+\rechts_K$,
  by~(\ref{E Path}), and by the definition of the formula 
  $\varphi_\leftrightarrow^\updownarrow$, 
  it follows directly that $(t,t')\in H$.

  Analogously, by applying~(\ref{E Northsouth}),
  for all $(n,m)\in\N\times\N$ 
  such that $\mu(n,m)=t$ and $\mu(n,m+1)=t'$, we conclude that
  $(t,t')\in V$.
\qed\end{pf}
Finally, we obtain the following theorem:

\begin{thm}
Satisfiability of \textnormal{PDL}$(k,n)$ is $\Sigma_1^1$-complete.
\end{thm}



\bibliographystyle{elsart-num}
\bibliography{bib.bib}

\end{document}